\documentclass[12pt]{elsart}

\usepackage{amssymb}
\usepackage[dvips]{graphicx}
\usepackage{latexsym}
\usepackage{amssymb}
\usepackage{cases}
\usepackage[compress]{cite}
\usepackage{xcolor}
\usepackage{enumerate}
\usepackage{mathrsfs}

\newtheorem{Theorem}{Theorem}

\journal{Physica A}

\begin{document}

\begin{frontmatter}

\title{Analysis of  diffusion and trapping efficiency for random walks on non-fractal scale-free trees}
\author[lable1,label2]{Junhao Peng}
\ead{pengjh@gzhu.edu.cn}
\author[label3]{Guoai Xu}
\address[lable1]{School of Mathematics and Information Science, Guangzhou University , Guangzhou 510006 ,  China.}
\address[label2]{Key Laboratory of Mathematics and Interdisciplinary Sciences of Guangdong
Higher Education Institutes, Guangzhou University,Guangzhou 510006 ,China.}
\address[label3]{State Key Laboratory of Networking and Switching Technology,
Beijing University of Posts and Telecommunications , Beijing 100876 ,China.}
\begin{abstract}
We study discrete random walks  on the NFSFT and  provide new methods to calculate the analytic  solutions of the MFPT for any pair of nodes, the MTT for any target node and MDT  for any source node.  Further more, using the MTT and the MDT as the measures of trapping efficiency and diffusion efficiency respectively,  we compare the trapping efficiency and diffusion efficiency for any two nodes  of NFSFT and find the best (or worst)  trapping sites  and the best (or worst) diffusion sites. Our results show that: the two hubs  of NFSFT is the best trapping site, but it is also the worst diffusion site, the   nodes which are the farthest nodes from the two hubs are the worst trapping sites, but they are also the best diffusion sites.  Comparing the maximum  and minimum of MTT and MDT, we found that the ratio between the maximum  and minimum  of MTT grows  logarithmically with network order, but the  ratio between the maximum  and minimum  of MTT  is almost equal to $1$. These results implie that  the trap's position has great effect on the trapping efficiency, but the position of source node almost has no effect on diffusion efficiency. We also conducted numerical simulation to test the results we have  derived, the results we derived are consistent with those obtained by numerical simulation.
\end{abstract}
\end{frontmatter}
\section{Introduction}
The problem of diffusion and trapping is part of the general problem for random walks. The range of applicability and of physical interest is enormous \cite{HaBe87, BuCa05, We94, LlMa01, MoTa11}.  Because many  materials encountered in nature exhibit fractal scaling\cite{SongHaMa05, SongHaMa06, RoHa07, Ta02} and
many problems in physics and chemistry are related to random walks on fractal structures \cite{Ko00, Avraham_Havlin04}, random walks on  fractal media  have attracted  a lot of interest in the past few years\cite{HaWe86, KaRe89, Mari89, MariSaSt93, RaTo83, BeTuKo10}. 
\par
The quantity we are interested in is the trapping time or mean first-passage time (MFPT), which is the expected number of steps to hit the target node(or trap) for the first time, for a walker starting from a source node.  Locating the target node at one special node and average the MFPTs over all the source nodes, we get mean trapping time(MTT) for the special node. Locating the source node at one special node and the average the MFPTs over all the target nodes,  we obtain mean diffusing time(MDT) for the special node. Both the MTT and MDT have different value for different nodes and they can be used as the measures of trapping efficiency and diffusion efficiency respectively. Comparing the  MTT and  MDT  among  all the network nodes, we can find the effects of node position on the trapping efficiency and diffusion efficiency. The nodes which  have the minimum MTT (or the maximum MTT) are best (or worst)  trapping sites and  the nodes which have the minimum MDT (or maximum MDT) are the best (or worst) diffusion sites .\par
 It is difficult to derive exact analytic solutions for MFPT on general fractal media, not to mention MTT and MDT. But  for deterministic fractals(or network), it  can be exactly studied. In the past several years, a lot of endeavors have been devoted to studying MFPT on different deterministic fractals(or networks)\cite{BeTuKo10, Mo69, GiMaNa94, KoBa02, MoHa89, BeMeTe08, HaRo08}. The MTT for some special nodes were  obtained for different deterministic fractals(or networks) such as Sierpinski  gaskets\cite{KoBa02}, Apollonian network\cite{ZhGuXi09}, pseudofractal scale-free web \cite{ZhQiZh09}, deterministic scale-free graph\cite{AgBuMa10} and some special trees\cite{CoMi10,  ZhZhGa10, LiZh13, LiWuZh11,  Agl08}. The MDT for some special nodes  were obtained for exponential treelike networks\cite{ZhLiLin11},  scale-free Koch networks\cite{ZhGa11} and deterministic scale-free graph\cite{AgBu09}. There were also some works focusing on global mean first-passage time (GMFPT), i.e., the average of MFPTs over all pairs of nodes, these results were  obtain for some special trees \cite{ZhZhGa10, LiZh13, LiWuZh11,  ZhWu10,  ZhYu09} and dual Sierpinski gaskets\cite{WuZh11}. \par
 However, the results of MTT and MDT which were obtained are only restricted to some special nodes for the above networks and we can not compare trapping efficiency and diffusing efficiency among all the network nodes.  It is still difficult to deriving the analytic solutions of  the MTT for any target node(or trap) and the MDT for any source node in these networks. It is also difficult to deriving the analytic solutions of MFPT for any pair of nodes.  \par
As for the recursive non-fractal scale-free trees(NFSFT),  the MTT for the hub node and the GMFPT had been obtained\cite {ZhLi11}. The MTT for some low-generation nodes can also be derived due to the methods of Ref. \cite{MeAgBeVo12}. But the  analytic calculations of MFPT for any pair of nodes, the MTT for any target node and the MDT for any source node were still unresolved.
\par
 In this paper,  we study unbiased discrete random walks  on the NFSFT, at each time step, the particle
(walker), starting from its current location, moves to any of its nearest neighbors with equal probability. Based on the self-similar structure of NFSFT and the relations between  random walks and electrical networks\cite{Te91, LO93},  we first provide new methods to derive analytic  solutions of the MFPT for any pair of nodes, the MTT for any target node and MDT  for any starting node, and then calculate the MTT and MDT for some special nodes of NFSFT, the result of MTT for the hubs  is consistent with those  derived in Ref. \cite{ZhLi11},  the other results which has never obtained in elsewhere   are consistent with those obtained by numerical simulation we conducted.  \par
 Further more, using the MTT and the MDT as the measures of trapping efficiency and diffusion efficiency respectively,  we compare the trapping efficiency and diffusion efficiency for any two nodes  of NFSFT and find the best ( or worst)  trapping sites  and the best (or worst) diffusing sites. Our results show that: the two hubs  of NFSFT is the best trapping site, but it is also the worst diffusing site, the   nodes which are the farthest nodes from the two hubs are the worst trapping sites, but they are also the best diffusion sites.  Comparing the maximum  and minimum of MTT and MDT, we found that the ratio between the maximum  and minimum  of MTT grows  logarithmically with network order, but the  ratio between the maximum  and minimum  of MTT  is almost equal to $1$. Thus the trap's position has great effect on the trapping efficiency, but the position of starting node almost has no effect on diffusion efficiency. The methods  we present can also be used  on other self-similar trees.
  \section{The network model and some notions}
\label{sec:1}
\begin{figure}
\begin{center}
\includegraphics[scale=0.3]{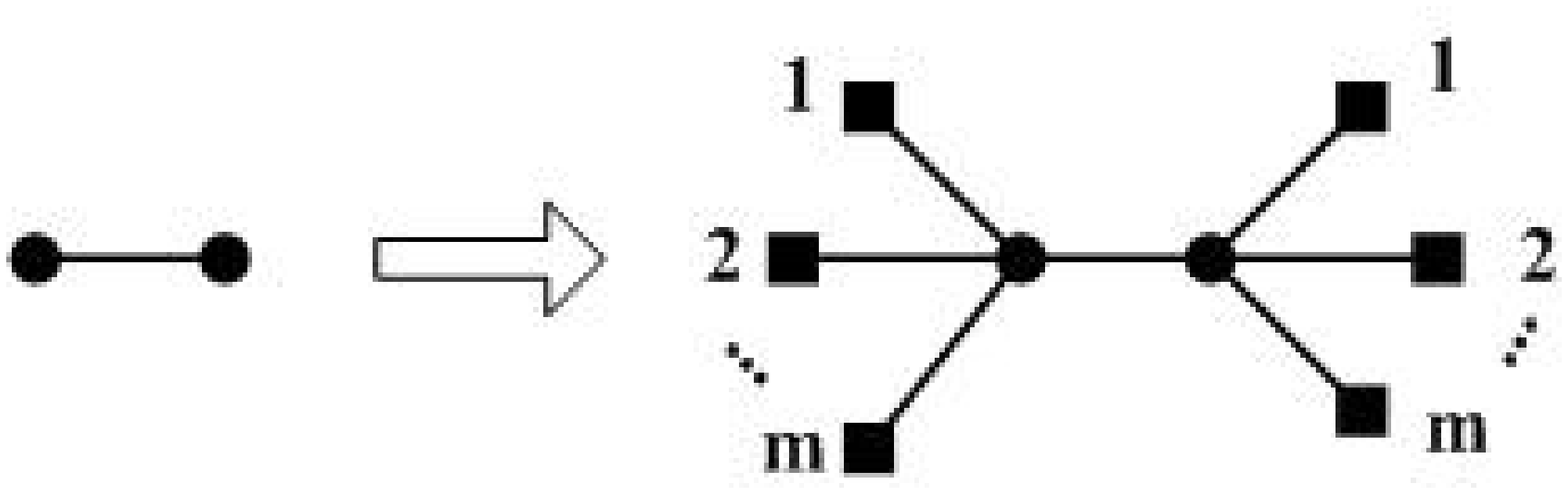}
\caption{Iterative construction method of the NFSFT}
\label{Edge_replace}       
\end{center}
\end{figure}

\begin{figure}
\begin{center}
\includegraphics[scale=0.3]{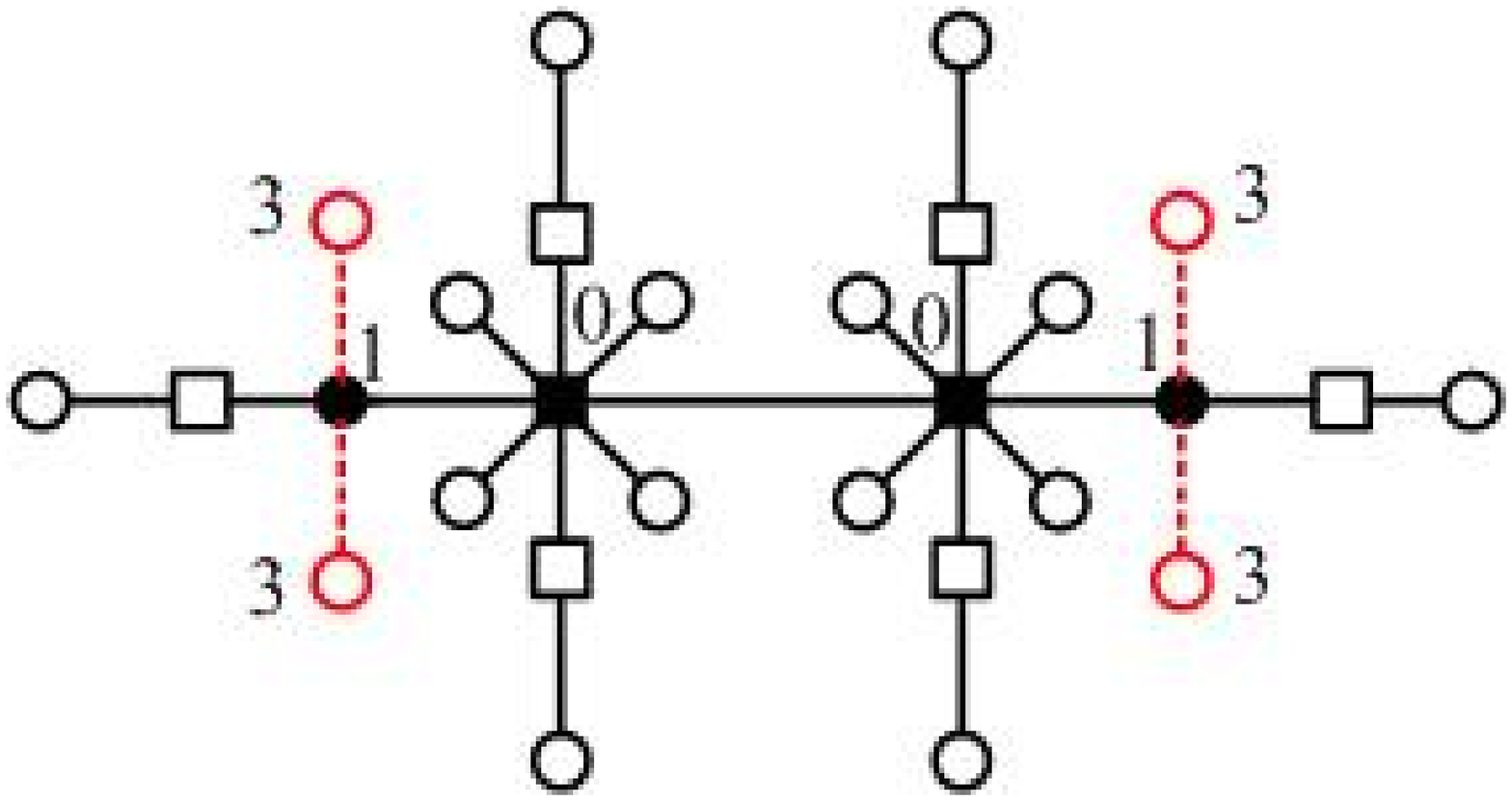}
\caption{The construction of the third generation NFSFT while  m = 1 and the Level information of its nodes:  solid square, level 0, solid  circle, level 1, hollow square, level 2; hollow circle, level 3. Four nodes represented by red hollow circle were all labeled as $\{0, 1, 3 \}$}
\label{level_nodes}
\end{center}
\end{figure}

The recursive non-fractal scale-free trees(NFSFT) we considered can be constructed iteratively\cite{JuKiKa02}. For convenience, we call the  times of iterations as the generation of the NFSFT and  denote by $G(t)$ the NFSFT of generation $t$. For $t = 0, G(0)$ is an edge connecting two nodes. For $t >0, G (t)$ is obtained from  $G(t-1)$ : for each of the existing edges  in $G(t-1)$, we introduce 2m (m is a positive integer) new nodes; half of them are connected to one end of the edge, and half of them are linked to the other end. That is, $G(t)$ is obtained from $G (t-1)$ via replacing every edge in $G(t-1)$ by the cluster on the right-hand side of the arrow in  Figure \ref{Edge_replace}. The construction of the third generation NFSFT for the particular case of m = 1 is shown in Figure  \ref{level_nodes}.\par
  The network family exhibits some striking properties of real-life systems, such as  scale free\cite{BaAl99, JuKiKa02} and  small-world properities\cite{Watts98, RoHa07, ZhangZhou07}.  In addition, they are non-fractal\cite{SongHaMa05, SongHaMa06, RoHa07}.
 According to its construction, one can easy obtain the total number of edges for $G(t)$ is $E_t=(2m+1)^t$ and the total number of nodes for $G(t)$ satisfies\cite{JuKiKa02, ZhLi11}
\begin{equation}
N_t=1+E_t=1+(2m+1)^t
\label{eq_nodes}
\end{equation}\par
 For convenience, we classify the nodes of $G(t)$ into different  levels. Nodes, which are generated during the $k$-th  iterations, are said to belong to level $k$ in this paper. For example,  in the third generation NFSFT with $m=1$, which  is shown in  Figure \ref{level_nodes}, the levels information of its nodes were shown as follows: nodes  represented by solid  square belong to level $0$.  Nodes represented by solid circle belong to level $1$. Nodes represented by hollow  square belong to level $2$. Nodes represented by hollow circle belong to level $3$. \par
 For any node $x$ of level $k$, there is a unique path $(V_0, V_1, V_2, ..., V_{n}, x ), (n\leq k-1)$ from the nearest node of level $0$ to node $x$. We call $\{V_0, V_1, V_2, ..., V_{n}\}$ the ancestors of node $x$ and $V_{n}$ the parent of node $x$. Thus the two nodes of level $0$ are the common ancestors of all other nodes, or all other nodes are the descendant nodes of the two nodes of level $0$.
 In this paper, we label the node of level $k$ by the sequence  $\{i_0, i_1, i_2, ..., i_{n}, k \}$, where $i_j$ is the level of node $V_j$, it is easy to know that $i_0=0$ and  $0< i_1< i_2< ...< i_{n}< k$ .  Although  different nodes may have the same labels,  nodes with the same label have the same properties base on the self-similar structure   of NFSFT. For example,  in the third generation NFSFT  shown in  Figure \ref{level_nodes}, the four nodes represented by red hollow circle were all labeled as $\{0, 1, 3 \}$. According to our method, for any node labeled as $\{i_0, i_1, i_2, ..., i_{n} \}$, its parent is labeled as $\{i_0, i_1, i_2, ..., i_{n-1} \}$, its ancestors are labeled as $\{i_0, i_1, i_2, ..., i_{k} \}(k=0, 1, \cdots, n-1)$.  \par
\begin{figure}
\begin{center}
\includegraphics[scale=0.3]{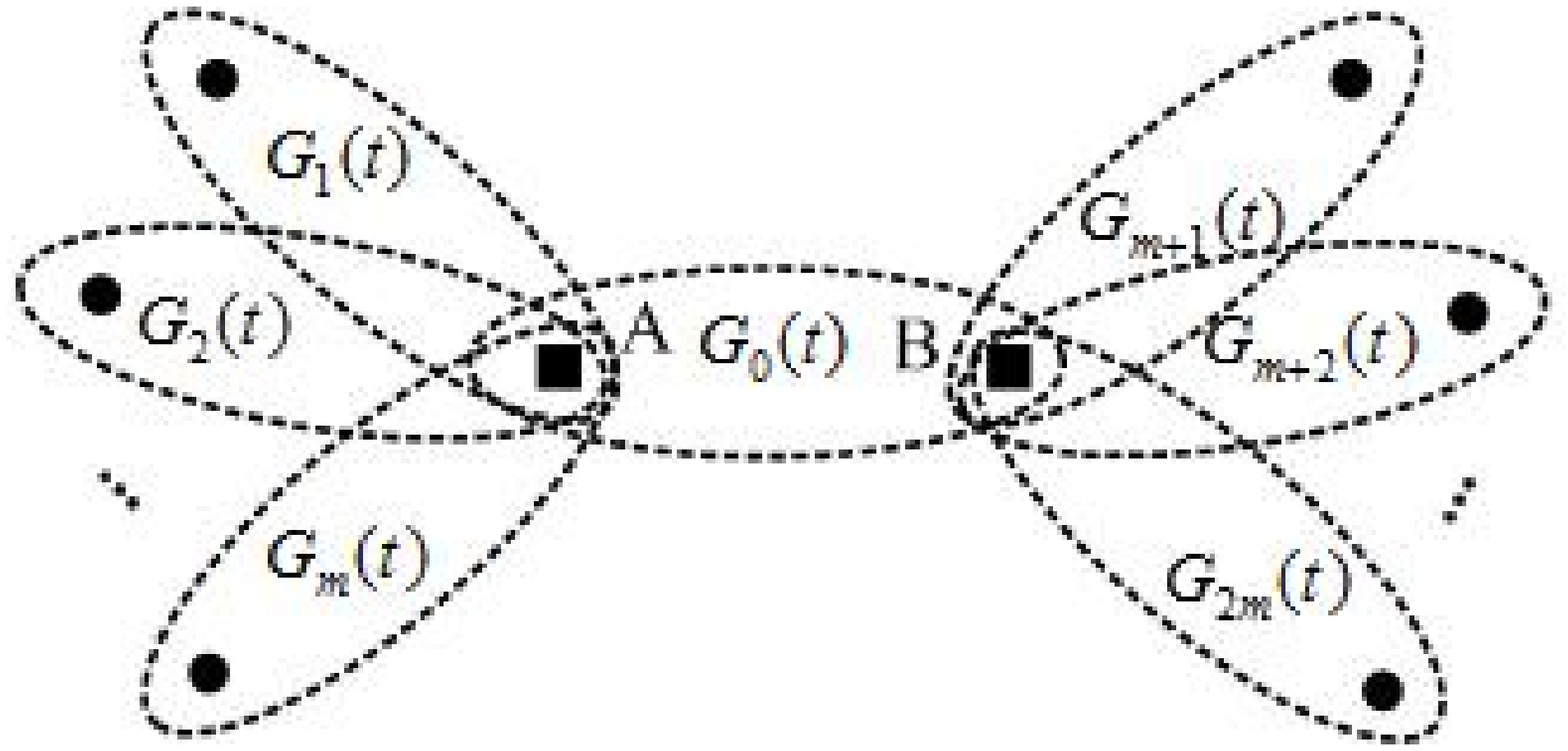}
\caption{Alternative construction of  NFSFT which highlights self-similarity: the NFSFT of generation $t$, denoted by $G(t)$, is composed of $2m+1$ copies of $G(t-1)$ which are labeled as  $G_0(t)$,  $G_1(t)$, $G_2(t)$, $\cdots$ ,$G_{2m}(t)$, and connected to one another at its two hubs $A$ and $B$.
}
\label{structure}
\end{center}
\end{figure}

The  NFSFT $G(t)$ can also be constructed by another method which is shown in Figure \ref{structure}: the NFSFT $G(t)$  is composed of $2m+1$ copies, called subunit, of $G(t-1)$ which are connected to one another at its two hubs (nodes with the highest degree).
 We also classify the subunits of  $G(t)$ into different levels and let $\Lambda_k$ denote the subunit of level $k(k\geq0)$. In this paper, $G(t)$ is said to be subunit of level $0$. For any   $k\geq0$, $\Lambda_k$ is composed of $2m+1$  subunits of level $k+1$. Thus, any edge of $G(t)$  is a subunit of level $t$ and $\Lambda_k$ is a copy of  NFSFT with generation $t-k$.
\section{Formulation of the problem}
\label{sec:gen_meth}
In this paper, we study  discrete-time random walks on FSFT $G(t)$. At each  step, the walker moves from its current location to any of its nearest neighbors with equal probability.  The quantity we are interested in is  mean first-passage time (MFPT), which is the expected number of steps to hit the target node(or trap) for the first time, for a walker starting from a source node.\par
Let $F(x,y)$  denote the MFPT from nodes $x$ to $y$ in NFSFT $G(t)$ and $\Omega$ denote the node set of $G(t)$, the sum
$$k(x,y)=F(x,y)+F(x,y)$$ is called the commute time and the MFPT can be expressed in term  of commute times\cite{Te91}.
\begin{equation}
F(x,y)=\frac{1}{2}\left(k(x,y)+\sum_{u\in \Omega}\pi(u)[k(y,u)-k(x,u)] \right)
\label{FXY}
\end{equation}
where  $\pi(u)=\frac{d_u}{2E_t}$ is the stationary distribution for random walks on the  NFSFT .\par
If we view the networks under consideration as electrical networks  by considering each edge to be a unit resistor and let $\Psi_{xy}$ denote the effective resistance  between two nodes $x$ and $y$ in the electrical networks,
 we have\cite{Te91}
\begin{equation}
k(x,y)=2E_t\Psi_{xy}
\label{KR}
\end{equation}\par
 where $E_t$ is the total numbers of edges of $G(t)$. Since the  NFSFT  we studied  are trees, the effective resistance between any two nodes is exactly the  shortest-path length  between the two nodes. Hence
 \begin{equation}
 \Psi_{xy}=L_{xy}
 \end{equation}
 where $L_{xy}$ denote  the shortest path length between node $x$ to node $y$.  Thus
 \begin{equation}
k(x,y)=2E_tL_{xy}
\label{KL}
\end{equation}
Substituting $k(x,y)$ with Eq.(\ref{KL})  in Eq.(\ref{FXY}), we obtain
\begin{eqnarray}
F(x,y)&=&E_t\left(L_{xy}+\sum_{u\in \Omega}\pi(u)L_{yu}-\sum_{u\in \Omega}\pi(u)L_{xu} \right)
\label{FXYL}
\end{eqnarray}\par
Thus we can derive the MFPT $F(x,y)$ for any two nodes $x$ and $y$ because we can calculate $\sum_{u\in G(t)}\pi(u)L_{xu}$ 
for any node $x$  in NFSFT. The detail methods will be shown in Sec.\ref{sec:det_meth}.\par
If we average the MFPTs over all the starting nodes and all target nodes, we  obtain MTT and MDT. That is to say, if we   define
\begin{eqnarray}
T_y&=&\frac{1}{E_t}\sum_{x\in \Omega,x\neq y}F(x,y) \label{MTTo}\\
D_x&=&\frac{1}{E_t}\sum_{y\in \Omega,y\neq x}F(x,y)  \label{MSTo}
\end{eqnarray}
$T_y$ is just the mean trapping time(MTT) for target node $y$ and $D_x$ is just mean diffusing time(MDT) for starting node $x$.
Let
\begin{equation}
S_x= \sum_{y \in \Omega}{L_{xy}}
\label{SX}
\end{equation}
\begin{equation}
W_x=\sum_{u\in \Omega}\pi(u)L_{xu}
\label{WY}
\end{equation}
\begin{equation}
\Sigma=\sum_{u\in \Omega}\left(\pi(u)\sum_{x\in \Omega}L_{xu}\right)
\label{WS}
\end{equation}
 Substituting  $F(x,y)$ with Eq.(\ref{FXYL}) in Eqs.(\ref{MTTo}) and (\ref{MSTo}), we obtain
\begin{eqnarray} \label{MTT}
T_y&=&\sum_{x\in \Omega,x\neq y}\left(L_{xy}+\sum_{u\in \Omega}\pi(u)L_{yu}-\sum_{u\in \Omega}\pi(u)L_{xu} \right)  \nonumber  \\
&=&\sum_{x\in \Omega,x\neq y}L_{xy}+\sum_{x\in \Omega,x\neq y}\sum_{u\in \Omega}\pi(u)L_{yu}-\sum_{x\in \Omega,x\neq y}\sum_{u\in \Omega}\pi(u)L_{xu} \nonumber  \\
&=&S_y+N_t\cdot W_y-\Sigma
\end{eqnarray}
\begin{eqnarray} \label{MDT}
D_x
   &=&S_x+\Sigma-N_t\cdot W_x
\end{eqnarray}
Hence, if we can calculate $\Sigma$ and $S_x, W_x$ for any node $x$, we can obtain MTT and MDT for any  node $x$. Although it is difficult to calculate these quantities for general tree, we presented methods for calculating these quantities for NFSFT based on its self-similar structure. Therefore, we can calculating  MTT and MDT for any  node. \par
\section{Methods for calculating MTT and MDT}
\label{sec:det_meth}
We first present detailed methods for calculating $S_x$, $\Sigma$ and $W_x$, and then calculating MFPT, MTT and MDT for some special nodes to explain our methods.
\subsection{Detailed methods for calculating $S_x$ and $W_x$ }
\label{SSLW}
According to the method in Sec.\ref{sec:1}, any node $x$ of NFSFT can be  labeled  by a sequence of nodes level information $\{0, i_1, i_2, ..., i_{n} \}$, $0< i_1< i_2< ...< i_{n}\leq t$. Although  different nodes may have the same labels,  nodes with the same label have the same $S_x$ and $W_x$  base on the Self-similar structure   of NFSFT. Thus we can use this sequence to represent  \textquotedblleft $x$\textquotedblright   in symbol \textquotedblleft $S_x$\textquotedblright and \textquotedblleft $W_x$\textquotedblright.  For example, for nodes $x$ of level $0$, they can be written as $S_{\{0\}}$ and $W_{\{0\}}$.  For the four nodes represented by red hollow circle, they can be written as $S_{\{0, 1, 3 \}}$ and $W_{\{0, 1, 3 \}}$.\par
First, we  calculate $S_{\{0\}}$ and $W_{\{0\}}$. 
 In order to tell the difference of $S_{\{0\}}$(and $W_{\{0\}}$)  for NFSFT of different generation $t$,
 let $S_{A}^t$, $W_{A}^t$ denote $S_{\{0\}}$ and $W_{\{0\}}$ in NFSFT of generation $t$ respectively.  It is easy to know $S_{A}^0=1$ and $W_{A}^0=\frac{1}{2}$. For $t>1$, according  to the self-similar structure shown in Figure \ref{structure}, $S_{A}^t$  satisfies the following recursion relation.
\begin{equation}
S_{A}^t=m\cdot S_{A}^{t-1}+S_{A}^{t-1} +m\cdot[S_{A}^{t-1}+(N_{t-1}-1)]  \nonumber
\end{equation}
For the right side of the equation, the first item represents the summation of shortest path length between node $A$ and nodes in the subunit $G_i(t)(i=1, 2, \cdots, m)$, the second item represents the summation of shortest path length between node $A$ and nodes in the subunit $G_0(t)$, the third item represents the summation of shortest path length between node $A$ and nodes in the subunit $G_i(t)(i=m+1, m+2, \cdots, 2m)$. Note that $N_{t-1}=(2m+1)^{t-1}+1$, thus, in NFSFT of generation $t$,
\begin{eqnarray}
S_{\{0\}} &=&S_{A}^t =(2m+1) S_{A}^{t-1}+m(2m+1)^{t-1}  \nonumber \\
      &=&(2m+1)\left[ (2m+1) S_{A}^{t-2}+m(2m+1)^{t-2}\right]+m(2m+1)^{t-1}  \nonumber \\
      &=&(2m+1)^2 S_{A}^{t-2}+2m(2m+1)^{t-1}  \nonumber \\
      &=&\cdots \nonumber \\
      &=&(2m+1)^t S_{A}^{0}+tm(2m+1)^{t-1}  \nonumber \\
      &=&(2m+1)^t+tm(2m+1)^{t-1}\label{SA}
\end{eqnarray} \par
Similarity
\begin{eqnarray}
W_{\{0\}}&=&W_{A}^t=\frac{m}{2m+1}\cdot W_{A}^{t-1}+\frac{1}{2m+1}W_{A}^{t-1} +\frac{m}{2m+1}\cdot[W_{A}^{t-1}+1]  \nonumber\\
             &=& W_{A}^{t-1}+\frac{m}{2m+1} \nonumber\\
             &=& W_{A}^{t-2}+\frac{2m}{2m+1} \nonumber\\
             &=&\cdots \nonumber \\
             &=&W_{A}^{0}+\frac{tm}{2m+1}  \nonumber \\
             &=&\frac{1}{2}+\frac{tm}{2m+1}
             \label{WA}
\end{eqnarray}\par
Now, we  calculate $S_{x}$ and $W_{x}$ for  node $x$ of any level. According to the method presented in Sec.\ref{sec:1}, $x$ can be labeled as  $\{0, i_1, i_2, ..., i_{n} \}$, $ 0< i_1< i_2< ...< i_{n}\leq t$, its parent, denoted by $p$, can only be labeled as $\{0, i_1, i_2, ..., i_{n-1} \}$.  We will derive the recursion relation  between $S_{\{0, i_1, i_2, ..., i_{n} \}}$ and $S_{\{0, i_1, i_2, ..., i_{n-1} \}}$.\par
Note that node  $x$ of level $i_n$ and its parent  $p$ are just two hubs of one subunit of level $i_{n}$ which is a copy of $G(t-i_n)$. The total numbers of nodes of this subunit is $N_{t-i_n}$, half of them are the descendant nodes of node $x$.  There is an edge between $x$ and $p$, node $x$ and its descendant nodes  connected with other nodes of the NFSFT by node $p$. Let $\Omega_{de}$ denote the set of the descendant nodes of node $x$, we have $\Omega=\Omega_{de}\bigcup \overline{\Omega}_{de}$. For any node $y\in \Omega_{de}$, $L_{xy}=L_{py}-1$, for any node $y\in \overline{\Omega}_{de}$, $L_{xy}=L_{py}+1$. Thus
\begin{eqnarray}
S_{\{0, i_1, i_2, ..., i_{n} \}}&=&S_x= \sum_{y \in \Omega}{L_{xy}}  \nonumber \\
&=&\sum_{y \in \Omega_{de}}{L_{xy}}+\sum_{y \in \overline{\Omega}_{de}}{L_{xy}}  \nonumber \\
&=&\sum_{y \in \Omega_{de}}{(L_{py}-1)}+\sum_{y \in \overline{\Omega}_{de}}{(L_{py}+1)} \nonumber \\
&=&\sum_{y \in \Omega_{de}}{L_{py}}-\frac{1}{2}N_{t-i_n}+\sum_{y \in \overline{\Omega}_{de}}{L_{py}}+N_t-\frac{1}{2}N_{t-i_n}  \nonumber \\
&=&\sum_{y \in \Omega}{L_{py}}+N_t-N_{t-i_n}  \nonumber \\
&=&S_p+N_t-N_{t-i_n}  \nonumber \\
&=&S_{\{0, i_1, i_2, ..., i_{n-1} \}}+(2m+1)^t-(2m+1)^{t-i_n}
\label{RSX}
\end{eqnarray}
Using Eq.(\ref{RSX}) repeatedly, we obtain
\begin{eqnarray}
S_{\{0, i_1, i_2, ..., i_{n} \}}&=&S_{\{0, i_1, i_2, ..., i_{n-1} \}}+(2m+1)^t-(2m+1)^{t-i_n}  \nonumber \\
&=&S_{\{0, i_1, i_2, ..., i_{n-2} \}}+2(2m+1)^t-(2m+1)^{t-i_{n-1}}-(2m+1)^{t-i_n}  \nonumber \\
&=&\cdots \nonumber \\
&=&S_{\{0\}}+n(2m+1)^t-\sum_{k=1}^{n}(2m+1)^{t-i_k}
\label{ESX}
\end{eqnarray}
Similarity
\begin{eqnarray}
W_{\{0, i_1, i_2, ..., i_{n} \}}&=&W_x= \sum_{y \in \Omega}{\pi(y)L_{xy}}  \nonumber \\
&=&\sum_{y \in \Omega_{de}}{\pi(y)L_{xy}}+\sum_{y \in \overline{\Omega}_{de}}{\pi(y)L_{xy}}  \nonumber \\
&=&\sum_{y \in \Omega_{de}}{\pi(y)(L_{py}-1)}+\sum_{y \in \overline{\Omega}_{de}}{\pi(y)(L_{py}+1)} \nonumber \\
&=&\sum_{y \in \Omega_{de}}{\pi(y)L_{py}}-\sum_{y \in \Omega_{de}}{\pi(y)}+\sum_{y \in \overline{\Omega}_{de}}{\pi(y)L_{py}}+\sum_{y \in \overline{\Omega}_{de}}{\pi(y)}  \nonumber \\
&=&\sum_{y \in \Omega}{\pi(y)L_{py}}-\frac{(2m+1)^{t-i_n}}{2(2m+1)^t}+ \frac{2(2m+1)^t-(2m+1)^{t-i_n}}{2(2m+1)^t} \nonumber \\
&=&W_p+ \frac{2(2m+1)^t-2(2m+1)^{t-i_n}}{2(2m+1)^t}  \nonumber \\
&=&W_{\{0, i_1, i_2, ..., i_{n-1} \}}+1-\frac{1}{(2m+1)^{i_n}}
\label{RWX}
\end{eqnarray}
Using Eq.(\ref{RWX}) repeatedly, we obtain
\begin{eqnarray}
W_{\{0, i_1, i_2, ..., i_{n} \}}&=&W_{\{0, i_1, i_2, ..., i_{n-1} \}}+1-\frac{1}{(2m+1)^{i_n}}  \nonumber \\
&=&W_{\{0, i_1, i_2, ..., i_{n-2} \}}+2-\frac{1}{(2m+1)^{i_{n-1}}}-\frac{1}{(2m+1)^{i_n}}  \nonumber \\
&=&\cdots \nonumber \\
&=&W_{\{0\}}+n-\sum_{k=1}^{n}\frac{1}{(2m+1)^{i_k}}
\label{EWX}
\end{eqnarray}
Thus, For any node $x$ labeled as  $\{0, i_1, i_2, ..., i_{n} \}$, we can exactly calculate $S_x$ and $W_x$ due to Eqs. (\ref{SA}), (\ref{WA}), (\ref{ESX}) and (\ref{EWX}).
 \subsection{Exact calculation of $\Sigma$ }
\label{SSWP}
Note that
 $$\Sigma=\sum_{u\in \Omega}(\pi(u)\sum_{x\in \Omega}L_{xu})=\frac{1}{2E_t}\sum_{u\in \Omega}(d_uS_u)$$
$\sum_{u\in \Omega}(d_uS_u)$ is just the summation of  $S_x$ for end nodes of any edges of $G(t)$.  For convenience, we label the two hubs of subunit $\Lambda_k$ as $A_k, B_k$. Because any edge of $G(t)$ is a subunit of level $t$,  its two end nodes is also its two hubs labeled as $A_t, B_t$.  Let
\begin{equation}
  \mathcal{S}^{(k)}\equiv \left( \begin{array}{c} S_{A_k}\\S_{B_k} \end{array} \right)
\end{equation}
We have
\begin{equation}
  \sum_{u\in \Omega}(d_uS_u)=\sum\left(\sum_{\Lambda_t}{\mathcal{S}^{(t)}}\right)
\end{equation}
For the right side of the equation, the second summation is run over all the subunits of  level $t$, the first summation is just add the two entries of $\sum_{\Lambda_t}{\mathcal{S}^{(t)}}$ together.\par
\begin{figure}
\begin{center}
\includegraphics[scale=0.5]{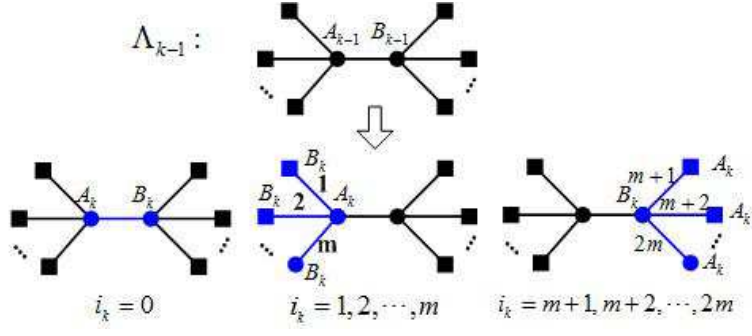}
\caption{Construction of $\Lambda_{k-1}$ and the relation  between the value of $i_k$ and the location of  subunit $\Lambda_{k}$ in $\Lambda_{k-1}$: subunit represented by blue line are the  subunit $\Lambda_{k}$ corresponding to value of $i_k$ below, whose two hubs are labeled as $A_k, B_k$. 
}
\label{lael_node}       
\end{center}
\end{figure}

 In order to calculate $\sum_{\Lambda_t}{\mathcal{S}^{(t)}}$, we label the subunit  $\Lambda_k$ by a sequence $\{i_1, i_2, ..., i_{k} \}$, where $i_j$ labels its position in the corresponding subunit $\Lambda_{j-1}$.  We assigning $i_k=0$ for the central one,  $i_k=1, 2, ..., m$ for the $m$ subunits containing hub $A_{k-1}$, $i_k=m+1, m+2, ..., 2m$ for the $m$ subunits containing hub $B_{k-1}$.  Figure \ref{lael_node} shows the construction of $\Lambda_{k-1}$ and the relation  between the value of $i_k$ and the location of  subunit $\Lambda_{k}$ in $\Lambda_{k-1}$: all subunit $\Lambda_{k}$ are represented by an edge, the one represented by blue edge are the  subunit $\Lambda_{k}$ corresponding to value of $i_k=0, 1, 2, \cdots, 2m+1$. We also build  mapping between hubs of $\Lambda_{k-1}$ and hubs of $\Lambda_{k}$:  hub labeled as $A_{k-1}$ in $\Lambda_{k-1}$ is also labeled as $A_k$ in $\Lambda_{k}$ while $i_k=0, 1, 2, \cdots, m$, hub labeled as $B_{k-1}$ in $\Lambda_{k-1}$ is also labeled as $B_k$ in $\Lambda_{k}$ while $i_k=0, m+1, m+2, \cdots, 2m$.
For example, for $i_k=1,2,\cdots, m$,  $A_{k}\equiv A_{k-1}$ and $A_{k-1}$ is parent of $B_{k}$. Note that  the label sequence of $B_{k}$ is ended with $k$(Because $B_{k}$ is a node of level $k$),  according to Eq.(\ref{RSX})
\begin{equation}
\left\{
\begin{array}{l}
S_{A_{k}}=S_{A_{k-1}}   \\
S_{B_{k}}=S_{A_{k-1}}+(2m+1)^t-(2m+1)^{t-k}
\end{array}
\right.
\label{RSAB}
\end{equation}
Eqs.(\ref{RSAB}) can also be rewritten as
\begin{equation}\label{SK1m}
  \mathcal{S}^{(k)}
=\left(                 
  \begin{array}{cc}
   1 & 0 \\
   1 & 0
  \end{array}
\right)
\mathcal{S}^{(k-1)}
+\left( \begin{array}{c} 0 \\(2m+1)^t-(2m+1)^{t-k} \end{array} \right)
\end{equation}
Similarly, one can define matrices $\mathcal{M}_{i_k}$ and $\mathcal{V}_{i_k}^k$ such that  equation (\ref{SK}) holds for $i_k=0, 1, \cdots, 2m$.
\begin{equation}\label{SK}
  \mathcal{S}^{(k)}=\mathcal{M}_{i_k}\mathcal{S}^{(k-1)}+\mathcal{V}_{i_k}^k
\end{equation}
with
\begin{equation} \label{M0}
\mathcal{M}_{0}=
\left(                 
  \begin{array}{cc}
   1 & 0 \\
   0 & 1
  \end{array}
\right),                  
\mathcal{V}_{0}^k=\left( \begin{array}{c} 0 \\ 0 \end{array} \right)
\end{equation}
\begin{equation} \label{M1m}
\mathcal{M}_{i}=
\left(                 
  \begin{array}{cc}
   1 & 0 \\
   1 & 0
  \end{array}
\right),   
\mathcal{V}_{i}^k=\left( \begin{array}{c} 0 \\(2m+1)^t-(2m+1)^{t-k} \end{array} \right),i=1,2,\cdots, m.
\end{equation}
and
\begin{equation} \label{Mm_2m}
\mathcal{M}_{i}=
\left(                 
  \begin{array}{cc}
   0 & 1 \\
   0 & 1
  \end{array}
\right),
\mathcal{V}_{i}^k=\left( \begin{array}{c} (2m+1)^t-(2m+1)^{t-k} \\0 \end{array} \right), i=m+1,\cdots, 2m.
\end{equation}
Using equation (\ref{SK}) repeatedly, we obtain
\begin{eqnarray}\label{SK0}
  \mathcal{S}^{(t)}&=&\mathcal{M}_{i_t}\mathcal{S}^{(t-1)}+\mathcal{V}_{i_t}^t  \nonumber \\
   &=&\mathcal{M}_{i_t}[\mathcal{M}_{i_{t-1}}\mathcal{S}^{(t-2)}+\mathcal{V}_{i_{t-1}}^{t-1}]+\mathcal{V}_{i_t}^t \nonumber \\
   &=& \mathcal{M}_{i_t}\mathcal{M}_{i_{t-1}}\mathcal{S}^{(t-2)}+\mathcal{M}_{i_t}\mathcal{V}_{i_{t-1}}^{t-1}+\mathcal{V}_{i_t}^t \nonumber \\
   &=& \cdots \nonumber \\
   &=&\mathcal{M}_{i_t}\mathcal{M}_{i_{t-1}}\cdots \mathcal{M}_{i_1}\mathcal{S}^{(0)}
   +\sum_{l=1}^{t-1}\mathcal{M}_{i_t}\mathcal{M}_{i_{t-1}}\cdots \mathcal{M}_{i_{l+1}}\mathcal{V}_{i_{l}}^{l}+\mathcal{V}_{i_t}^t
\end{eqnarray}
where
\begin{equation}\label{S0}
  \mathcal{S}^{(0)}\equiv \left( \begin{array}{c} S_{A_0}\\S_{B_0} \end{array} \right)=S_{\{0\}} \left( \begin{array}{c} 1\\1 \end{array} \right)
\end{equation}\par
Because any  subunit of level $t$  is in one to one correspondence with a path  $\{i_1,\cdots,i_t\}$, let $\{i_1,\cdots,i_t\}$ run over all the possible values and calcute $\sum_{\{i_1,\cdots,i_t\}} \mathcal{S}^{(t)}$, the summation of the two entries of $\sum_{\{i_1,\cdots,i_t\}} \mathcal{S}^{(t)}$ is just equal to $\sum_{u\in \Omega}(d_uS_u)$.  
 Making use of the following identity
\[
\sum_{\{i_1,\cdots,i_t\}} \sum_{l=1}^{t-1} = \sum_{l=1}^{t-1} \sum_{\{i_1,\cdots,i_t\}},
\]
and define
\begin{equation} \label{MTotal}
\mathcal{M}_{tot} =\sum_{i=0}^{2m} \mathcal{M}_{i}
\end{equation}
\begin{equation} \label{VTotal}
\mathcal{V}_{tot}^l = \sum_{i=0}^{2m} \mathcal{V}_{i}^{l}
\end{equation}
 we have
\begin{eqnarray}\label{SuoV}
  \sum_{\{i_1,\cdots,i_t\}}\mathcal{M}_{i_t}\mathcal{M}_{i_{t-1}}\cdots \mathcal{M}_{i_{l+1}}\mathcal{V}_{i_{l}}^{l}= (2m+1)^{l-1}\mathcal{M}_{tot}^{t-l}\mathcal{V}_{tot}^l
\end{eqnarray}
Thus
\begin{eqnarray}\label{SST}
  \sum_{\{i_1,\cdots,i_t\}} \mathcal{S}^{(t)}&=& \sum_{\{i_1,\cdots,i_t\}}\left[ \mathcal{M}_{i_t}\mathcal{M}_{i_{t-1}}\cdots \mathcal{M}_{i_1}\mathcal{S}^{(0)} \right. \nonumber \\
    &&\left.+\sum_{l=1}^{t-1}\mathcal{M}_{i_t}\mathcal{M}_{i_{t-1}}\cdots \mathcal{M}_{i_{l+1}}\mathcal{V}_{i_{l}}^{l}+\mathcal{V}_{i_t}^t \right] \nonumber \\
   &=&\mathcal{M}_{tot}^{t}\mathcal{S}^{(0)}+\sum_{l=1}^{t-1}(2m+1)^{l-1}\mathcal{M}_{tot}^{t-l}\mathcal{V}_{tot}^l+(2m+1)^{t-1}\mathcal{V}_{tot}^t  \nonumber \\
   &=& \mathcal{M}_{tot}^{t}\mathcal{S}^{(0)}+\sum_{l=1}^{t}(2m+1)^{l-1}\mathcal{M}_{tot}^{t-l}\mathcal{V}_{tot}^l
\end{eqnarray}
 Substituting  $\mathcal{M}_{i}$ with Eq.(\ref{M0}), (\ref{M1m}) and (\ref{Mm_2m}) in Eq. (\ref{MTotal}), and orthogonal decomposing $\mathcal{M}_{tot}$, we obtain
 \begin{equation} \label{MTD}
\mathcal{M}_{total}=
\left(                 
  \begin{array}{cc}
   m+1 & m \\
   m & m+1
  \end{array}
\right)=                  
\left(                 
  \begin{array}{cc}
\frac{\sqrt{2}}{2} & -\frac{\sqrt{2}}{2} \\
   \frac{\sqrt{2}}{2} & \frac{\sqrt{2}}{2}
  \end{array}
\right)                 
\left(                 
  \begin{array}{cc}
   2m+1 & 0 \\
   0 & 1
  \end{array}
\right)                 
\left(                 
  \begin{array}{cc}
\frac{\sqrt{2}}{2} & \frac{\sqrt{2}}{2} \\
   -\frac{\sqrt{2}}{2} & \frac{\sqrt{2}}{2}
  \end{array}
\right)                 
\end{equation}
Therefore,
\begin{equation} \label{MTK}
\mathcal{M}_{total}^k=
\left(                 
  \begin{array}{cc}
\frac{\sqrt{2}}{2} & -\frac{\sqrt{2}}{2} \\
   \frac{\sqrt{2}}{2} & \frac{\sqrt{2}}{2}
  \end{array}
\right)                 
\left(                 
  \begin{array}{cc}
   (2m+1)^k & 0 \\
   0 & 1
  \end{array}
\right)                 
\left(                 
  \begin{array}{cc}
\frac{\sqrt{2}}{2} & \frac{\sqrt{2}}{2} \\
   -\frac{\sqrt{2}}{2} & \frac{\sqrt{2}}{2}
  \end{array}
\right)                 
\end{equation}
 Substituting  $\mathcal{V}_{i}$ with with Eq.(\ref{M0}), (\ref{M1m}) and (\ref{Mm_2m}) in Eq. (\ref{VTotal}), we get
 \begin{equation} \label{VT}
\mathcal{V}_{tot}^l=m\left[(2m+1)^t-(2m+1)^{t-l}\right]\left( \begin{array}{c} 1 \\ 1 \end{array} \right)
\end{equation}
 Thus
 \begin{eqnarray}\label{MTS0}
  \mathcal{M}_{tot}^{t}\mathcal{S}^{(0)}&=&
  \left(                 
  \begin{array}{cc}
\frac{\sqrt{2}}{2} & -\frac{\sqrt{2}}{2} \\
   \frac{\sqrt{2}}{2} & \frac{\sqrt{2}}{2}
  \end{array}
\right)                 
\left(                 
  \begin{array}{cc}
   (2m+1)^t & 0 \\
   0 & 1
  \end{array}
\right)                 
\left(                 
  \begin{array}{cc}
\frac{\sqrt{2}}{2} & \frac{\sqrt{2}}{2} \\
   -\frac{\sqrt{2}}{2} & \frac{\sqrt{2}}{2}
  \end{array}
\right)                 
\left( \begin{array}{c} 1\\1\end{array} \right)S_{\{0\}}
 \nonumber \\
   &=&\left(                 
  \begin{array}{cc}
\frac{\sqrt{2}}{2} & -\frac{\sqrt{2}}{2} \\
   \frac{\sqrt{2}}{2} & \frac{\sqrt{2}}{2}
  \end{array}
\right)                 
\left(                 
  \begin{array}{cc}
   (2m+1)^t & 0 \\
   0 & 1
  \end{array}
\right)                 
 \left( \begin{array}{c} \sqrt{2}\\0 \end{array} \right) S_{\{0\}}
\nonumber \\
   &=&\left(                 
  \begin{array}{cc}
\frac{\sqrt{2}}{2} & -\frac{\sqrt{2}}{2} \\
   \frac{\sqrt{2}}{2} & \frac{\sqrt{2}}{2}
  \end{array}
\right)                 
 \left( \begin{array}{c} (2m+1)^t \sqrt{2}\\0 \end{array} \right) S_{\{0\}} \nonumber \\
   &=& (2m+1)^t S_{\{0\}}
     \left( \begin{array}{c} 1\\ 1 \end{array} \right)
\end{eqnarray}
\begin{eqnarray}\label{MTVT}
  &&\sum_{l=1}^{t}(2m+1)^{l-1}\mathcal{M}_{tot}^{t-l}\mathcal{V}_{tot}^l\nonumber \\
  &=&   \sum_{l=1}^{t}(2m+1)^{l-1}
  \left\{
    \left(                 
  \begin{array}{cc}
\frac{\sqrt{2}}{2} & -\frac{\sqrt{2}}{2} \\
   \frac{\sqrt{2}}{2} & \frac{\sqrt{2}}{2}
  \end{array}
\right)                 
\left(                 
  \begin{array}{cc}
   (2m+1)^{t-l} & 0 \\
   0 & 1
  \end{array}
\right)                 
  \right.       \nonumber \\         
&&\left.
\left(                 
  \begin{array}{cc}
\frac{\sqrt{2}}{2} & \frac{\sqrt{2}}{2} \\
   -\frac{\sqrt{2}}{2} & \frac{\sqrt{2}}{2}
  \end{array}
\right)\cdot           
m\left[(2m+1)^t-(2m+1)^{t-l}\right]\left( \begin{array}{c} 1 \\ 1 \end{array} \right)\right\}
 \nonumber \\
  &=&   \sum_{l=1}^{t}m(2m+1)^{l-1}\left[(2m+1)^t-(2m+1)^{t-l}\right](2m+1)^{t-l}\left( \begin{array}{c} 1 \\ 1 \end{array} \right)
  \nonumber \\
  &=&   \sum_{l=1}^{t}m(2m+1)^{t-1}\left[(2m+1)^t-(2m+1)^{t-l}\right]\left( \begin{array}{c} 1 \\ 1 \end{array} \right)
 \nonumber \\
  &=&\left[tm(2m+1)^{2t-1}-\sum_{l=1}^{t}m(2m+1)^{2t-l-1}\right]\left( \begin{array}{c} 1 \\ 1 \end{array} \right) \nonumber \\
  &=&  \left[tm(2m+1)^{2t-1}-\frac{1}{2}(2m+1)^{2t-1}+\frac{1}{2}(2m+1)^{t-1}\right]\left( \begin{array}{c} 1 \\ 1 \end{array} \right)
\end{eqnarray}
Inserting Eqs. (\ref{S0}), (\ref{MTS0}), (\ref{MTVT}) into Eq.(\ref{SST}), calculating the summation of the two entries of $\sum_{\{i_1,\cdots,i_t\}} \mathcal{S}^{(t)}$,  and denoting the summation by $Sum$, we obtain
\begin{eqnarray}\label{SS}
  Sum&=&2 \left[tm(2m+1)^{2t-1}-\frac{1}{2}(2m+1)^{2t-1}+\frac{1}{2}(2m+1)^{t-1}+ (2m+1)^t S_{\{0\}}\right]\nonumber \\
   &=& 4tm(2m+1)^{2t-1}+2(2m+1)^{2t}-(2m+1)^{2t-1}+(2m+1)^{t-1}
\end{eqnarray}
Since  $\sum_{u\in \Omega}(d_uS_u)=Sum$ and $E_t=(2m+1)^t$, therefore,
\begin{eqnarray}\label{SAPL}
\Sigma
&=&\frac{1}{2E_t}\sum_{u\in \Omega}\left(d_u S_u\right)\nonumber \\
&=&2tm(2m+1)^{t-1}+(2m+1)^{t}-\frac{1}{2}(2m+1)^{t-1}+\frac{1}{2(2m+1)}
\end{eqnarray}
\subsection{Examples}
\label{sec:example}
According to the methods presented in Sec.\ref{SSLW}  and Sec.\ref{SSWP},  we can calculate $\Sigma$ and $S_x,  W_x$ for any node $x$ of $G(t)$.
We don't intend to calculate these quantities for every  node of $G(t)$ because the total number of nodes increasing rapidly with the growth of $t$.
As shown in Sec.\ref{sec:1}, any node $x$ of NFSFT can be  labeled  by a sequence of nodes level information $\{0, i_1, i_2, ..., i_{n} \}$, $0< i_1< i_2< ...< i_{n}\leq t$. In order to explain our methods,   we calculate the  MTT or MDT for nodes of level $0$  labeled as $\{0\}$ (i.e., $A, B$ in Figure  \ref{structure}) and nodes of level $k$ labeled as $\{0, 1, 2,\cdots, k\}(1\leq k\leq t)$, which are the farthest nodes from  node labeled as {\{0\}} among all nodes of level $k$. Similar to Sec.\ref{SSLW}, we  use the label sequence to represent  $x$ in symbol \textquotedblleft $T_x$ \textquotedblright and \textquotedblleft $D_x$ \textquotedblright.\par
For nodes  of level $0$, 
inserting Eqs.(\ref{SA} ), (\ref{WA} ) and (\ref {SAPL} )  into Eq.(\ref{MTT}) and Eq.(\ref{MDT}), we obtain the MTT  and MDT for nodes labeled as $\{0\}$.
\begin{eqnarray} \label{MTT0}
T_{\{0\}}
&=&S_{\{0\}}+N_t W_{\{0\}}-\Sigma  \nonumber  \\
&=&(m+1)(2m+1)^{t-1}+\frac{m(t+1)}{2(2m+1)}
\end{eqnarray}
and
\begin{eqnarray} \label{MDT0}
D_{\{0\}}  &=&S_{\{0\}}+\Sigma-N_t W_{\{0\}}  \nonumber  \\
   &=&\frac{3m+1}{2m+1}(2m+1)^{t}+2tm(2m+1)^{t-1}-\frac{m(t+1)}{2(2m+1)}
\end{eqnarray}
These result of $T_{\{0\}}$   is consistent with those  derived in Ref. \cite{ZhLi11}. \par
For nodes of level $k(1\leq k\leq t)$, we only study the nodes labeled as $\{0, 1, 2,\cdots, k\}$, which are the farthest nodes from  node labeled as {\{0\}} among all nodes of level $k$.
According to Eqs.(\ref{ESX}) and (\ref{EWX}), we get
\begin{eqnarray}
& &S_{\{0, 1, 2,\cdots, k\}}\nonumber \\
&=&S_{\{0\}}+k(2m+1)^t-\sum_{i=1}^{k}(2m+1)^{t-i}  \nonumber \\
&=&S_{\{0\}}+k(2m+1)^t-\frac{(2m+1)^t}{2m}+\frac{(2m+1)^{t-k}}{2m}   \nonumber \\
&=&tm(2m+1)^{t-1}+(k+1)(2m+1)^t-\frac{(2m+1)^t}{2m}+\frac{(2m+1)^{t-k}}{2m}
\label{ES0_n}
\end{eqnarray}
and
\begin{eqnarray}
W_{\{0, 1, 2,\cdots, k\}}&=&W_{\{0\}}+k-\sum_{i=1}^{k}\frac{1}{(2m+1)^{i}}  \nonumber \\
&=&W_{\{0\}}+k-\frac{1}{2m}+\frac{1}{2m(2m+1)^{k}}  \nonumber \\
&=&\frac{tm}{2m+1}+k+\frac{m-1}{2m}+\frac{1}{2m(2m+1)^{k}}
\label{EW0_n}
\end{eqnarray}
Thus
\begin{eqnarray}
N_tW_{\{0, 1, 2,\cdots, k\}}&=&[(2m+1)^{t}+1]W_{\{0\}}  \nonumber \\
&=&(k+\frac{m-1}{2m})(2m+1)^{t}+tm(2m+1)^{t-1}+\frac{(2m+1)^{t-k}}{2m}    \nonumber \\
&&+\frac{tm}{2m+1}+k+\frac{m-1}{2m}+\frac{1}{2m(2m+1)^{k}}
\label{EW0_n}
\end{eqnarray}
Therefore
\begin{eqnarray} \label{MTT0_n}
T_{\{0, 1, 2,\cdots, k\}}
&=&S_{\{0, 1, 2,\cdots, k\}}+N_t W_{\{0, 1, 2,\cdots, k\}}-\Sigma  \nonumber  \\
&=&2(k-\frac{1}{2m})(2m+1)^t+(m+1)(2m+1)^{t-1}+\frac{(2m+1)^{t-k}}{m} \nonumber  \\
&&+\frac{1}{2m(2m+1)^{k}}+k+\frac{2tm^2+2m^2-2m-1}{2m(2m+1)}
\end{eqnarray}
and
\begin{eqnarray} \label{MDT0_n}
D_{\{0, 1, 2,\cdots, k\}}  &=&S_{\{0, 1, 2,\cdots, k\}}+\Sigma-N_t W_{\{0, 1, 2,\cdots, k\}}  \nonumber  \\
   &=&2(2m+1)^t+(2tm-m-1)(2m+1)^{t-1} \nonumber  \\
&&-\frac{1}{2m(2m+1)^{k}}-k-\frac{2tm^2+2m^2-2m-1}{2m(2m+1)}
\end{eqnarray}
We also conducted numerical simulation to test the results we have just derived, the results just derived are consistent with those obtained by numerical simulation.
\section{Analysis of trapping efficiency and diffusion efficiency for random walks on NFSFT}
\label{sec:Comparison and analysis}
Using the MTT and the MDT as the measure of trapping efficiency and diffusion efficiency respectively, we compare the trapping efficiency and diffusion efficiency for any two  nodes of NFSFT and obtain the following results.
\begin{Theorem} \label{TheMTT_MDT}  
 For any two nodes $x$ and $y$ of NFSFT, They can be labeled as $\{0, i_1, i_2, ..., i_{n_x} \}$ and $\{0, j_1, j_2, ..., j_{n_y} \}$ respectively,
  \begin{itemize}
\item  if $n_x>n_y$,  we have \\
$T_{\{0, i_1, i_2, ..., i_{n_x} \}}> T_{\{0, j_1, j_2, ..., j_{n_y} \}}$ and $D_{\{0, i_1, i_2, ..., i_{n_x} \}}< D_{\{0, j_1, j_2, ..., j_{n_y} \}}$
\item if $n_x<n_y$,  we have \\
$T_{\{0, i_1, i_2, ..., i_{n_x} \}}<T_{\{0, j_1, j_2, ..., j_{n_y} \}}$ and $D_{\{0, i_1, i_2, ..., i_{n_x} \}}> D_{\{0, j_1, j_2, ..., j_{n_y} \}}$
 \item if $n_x=n_y$,
           \begin{enumerate}
            \item there is a positive integer $k(1\leq k \leq n_x)$, such that $i_l=j_l$ holds for $l=1,2,\cdots,k-1$, but $i_k\neq j_k$,\begin{enumerate}
                 \item  if $i_k< j_k$, we have \\
                 $T_{\{0, i_1, i_2, ..., i_{n_x} \}}<T_{\{0, j_1, j_2, ..., j_{n_y} \}}$ and $D_{\{0, i_1, i_2, ..., i_{n_x} \}}> D_{\{0, j_1, j_2, ..., j_{n_y} \}}$
                 \item  if $i_k>j_k$, we have \\
                $T_{\{0, i_1, i_2, ..., i_{n_x} \}}>T_{\{0, j_1, j_2, ..., j_{n_y} \}}$ and $D_{\{0, i_1, i_2, ..., i_{n_x} \}}< D_{\{0, j_1, j_2, ..., j_{n_y} \}}$
             \end{enumerate}
            \item $i_l=j_l$ holds for $l=1,2,\cdots, n_x$,  we have \\
            $T_{\{0, i_1, i_2, ..., i_{n_x} \}}= T_{\{0, j_1, j_2, ..., j_{n_y} \}}$
            \end{enumerate}
\end{itemize}
\end{Theorem}
The proof of {\bf Theorem.\ref{TheMTT_MDT}} was provided in Sec.\ref{sec:Proof}. Using {\bf Theorem.\ref{TheMTT_MDT}},  we found
$$T_{\{0 \}}<T_{\{0, 1 \}}<T_{\{0, 2 \}}<\cdots<T_{\{0, 1, 2 \}}<\cdots<T_{\{0, t-1, t \}}<\cdots<T_{\{0, 1, 2, ..., t \}}$$
and
$$D_{\{0 \}}>D_{\{0, 1 \}}>D_{\{0, 2 \}}>\cdots>D_{\{0, 1, 2 \}}>\cdots>D_{\{0, t-1, t \}}>\cdots>D_{\{0, 1, 2, ..., t \}}$$
 Results shows: nodes  labeled as ${\{0 \}}$ which is the two hubs of NFSFT, have minimum MTT and  maximum MDT, hence they are  the best trapping site  and worst diffusion site. Nodes labeled as ${\{0, 1, 2, ..., t \}}$, which is  the farthest nodes from hubs, have maximum MTT and minimum MDT, therefore they are the worst trapping sites  and best diffusing sites.\par
Let $k=t$ in Eqs.  (\ref{MTT0_n}) and (\ref{MDT0_n}),we obtain $T_{\{0, 1, 2, ..., t \}}$ and $D_{\{0, 1, 2, ..., t \}}$.
Comparing $T_{\{0, 1, 2, ..., t \}}$ with $T_{\{0 \}}$ shown in Eq. (\ref{MTT0}), we have 
 \begin{equation} \label{Com_MTT}
\frac{T_{\{0, 1, 2, ..., t \}}}{ T_{\{0 \}}}\approx 2(t-\frac{1}{2m})\frac{2m+1}{m+1}+1\propto log_{2m+1}N_t
\end{equation}
where $N_t$ is the total number of nodes for NFSFT. Eq.(\ref{Com_MTT}) shows that the ratio between the maximum  and minimum  of MTT grows  logarithmically with network order , thus the trap's position has great effect on the trapping efficiency.\par
Comparing $D_{\{0, 1, 2, ..., t \}}$ with $D_{\{0 \}}$ shown in Eq. (\ref{MDT0}), we obtain
 \begin{equation} \label{Com_MDT}
\frac{D_{\{0, 1, 2, ..., t \}}}{ D_{\{0 \}}}\approx 1-\frac{t}{(2m+1)^{t-1}(2tm+3m+1)}\approx 1
\end{equation}
which shows that the difference between maximum  and minimum of MDT is quite small, thus the position of starting node almost has no effect on diffusion efficiency.

\section{Conclusion}
\label{sec:4}
In this paper,we study unbiased discrete random walks on NFSFT. First, we  provided general methods for calculating  the mean trapping time(MTT) for any target node and the mean diffusing time(MDT)  for any source node, and then we gave some examples to explain our methods.
Finally, using the MTT and the MDT as the measures of trapping efficiency and diffusion efficiency respectively, we compare the trapping efficiency and diffusion efficiency for any two nodes  of NFSFT and find the best ( or worst)  trapping sites  and the best ( or worst) diffusing sites. Our results show that: the two hubs  of NFSFT is the best trapping site, but it is also the worst diffusing site, the   nodes which are the farthest nodes from the two hubs are the worst trapping sites, but they are also the best diffusion sites.  Comparing the maximum  and minimum of MTT and MDT, we found that the  maximum  and minimum of MTT have great difference, but the difference between maximum  and minimum of MDT is quite small, thus the trap's position has great effect on the trapping efficiency, but the position of starting node almost has no effect on diffusion efficiency. The methods  we present can also be used  on other self-similar trees.\par

\section*{Acknowledgment}

The authors are grateful to the anonymous referees for their valuable comments and suggestions. This work was supported  by
the scientific research program of Guangzhou municipal colleges and universities under Grant No. 2012A022.

\appendix 
\section{Proof of { Theorem.\ref{TheMTT_MDT}} }
\label{sec:Proof}

For any two nodes $x$ and $y$ labeled as $\{0, i_1, i_2, ..., i_{n_x} \}$ and $\{0, j_1, j_2, ..., j_{n_y} \}$ respectively, we have
According to Eqs.(\ref{MTT}), (\ref{ESX}), (\ref {EWX}), we have
\begin{eqnarray}\label{TX_TY}
&&T_{\{0, i_1, i_2, ..., i_{n_x} \}}- T_{\{0, j_1, j_2, ..., j_{n_y} \}} \nonumber\\
&=&S_{\{0, i_1, i_2, ..., i_{n_x} \}}+N_t W_{\{0, i_1, i_2, ..., i_{n_x} \}}-\left[S_{\{0, i_1, i_2, ..., i_{n_y} \}}+N_tW_{\{0, i_1, i_2, ..., i_{n_y} \}}\right] \nonumber\\
&=&S_{\{0, i_1, i_2, ..., i_{n_x} \}}-S_{\{0, i_1, i_2, ..., i_{n_y} \}}+N_t\left[W_{\{0, i_1, i_2, ..., i_{n_x} \}}-W_{\{0, i_1, i_2, ..., i_{n_y} \}}\right] \nonumber\\
&=&(n_x-n_y)(2m+1)^t-\sum_{k=1}^{n_x}(2m+1)^{t-i_k}+\sum_{k=1}^{n_y}(2m+1)^{t-j_k} \nonumber\\
&&+\left[(2m+1)^t+1\right]\left[(n_x-n_y)-\sum_{k=1}^{n_x}(2m+1)^{-i_k}+\sum_{k=1}^{n_y}(2m+1)^{-j_k}\right] \nonumber\\
&=&2(n_x-n_y)(2m+1)^t-2\sum_{k=1}^{n_x}(2m+1)^{t-i_k}+2\sum_{k=1}^{n_y}(2m+1)^{t-j_k} \nonumber\\
&&+(n_x-n_y)-\sum_{k=1}^{n_x}(2m+1)^{-i_k}+\sum_{k=1}^{n_y}(2m+1)^{-j_k}
\end{eqnarray}
and
\begin{eqnarray}\label{DX_DY}
&&D_{\{0, i_1, i_2, ..., i_{n_x} \}}- D_{\{0, j_1, j_2, ..., j_{n_y} \}} \nonumber\\
&=&S_{\{0, i_1, i_2, ..., i_{n_x} \}}-N_t W_{\{0, i_1, i_2, ..., i_{n_x} \}}-\left[S_{\{0, i_1, i_2, ..., i_{n_y} \}}-N_tW_{\{0, i_1, i_2, ..., i_{n_y} \}}\right] \nonumber\\
&=&S_{\{0, i_1, i_2, ..., i_{n_x} \}}-S_{\{0, i_1, i_2, ..., i_{n_y} \}}-N_t\left[W_{\{0, i_1, i_2, ..., i_{n_x} \}}-W_{\{0, i_1, i_2, ..., i_{n_y} \}}\right] \nonumber\\
&=&(n_x-n_y)(2m+1)^t-\sum_{k=1}^{n_x}(2m+1)^{t-i_k}+\sum_{k=1}^{n_y}(2m+1)^{t-j_k} \nonumber\\
&&-\left[(2m+1)^t+1\right]\left[(n_x-n_y)-\sum_{k=1}^{n_x}(2m+1)^{-i_k}+\sum_{k=1}^{n_y}(2m+1)^{-j_k}\right] \nonumber\\
&=&(n_y-n_x)+\sum_{k=1}^{n_x}(2m+1)^{-i_k}-\sum_{k=1}^{n_y}(2m+1)^{-j_k}
\end{eqnarray}
The result of Eqs.(\ref{TX_TY})and (\ref{DX_DY}) can be divided into $3$ case.\par
Case I. If $n_x>n_y$, then $n_x-n_y\geq1$. Note that $t\geq n_x>n_y\geq 1$, we obtain
\begin{eqnarray}\label{TX_TY1}
&&T_{\{0, i_1, i_2, ..., i_{n_x} \}}- T_{\{0, j_1, j_2, ..., j_{n_y} \}} \nonumber\\
&=&2(n_x-n_y)(2m+1)^t-2\sum_{k=1}^{n_x}(2m+1)^{t-i_k}+2\sum_{k=1}^{n_y}(2m+1)^{t-j_k} \nonumber\\
&&+(n_x-n_y)-\sum_{k=1}^{n_x}(2m+1)^{-i_k}+\sum_{k=1}^{n_y}(2m+1)^{-j_k} \nonumber\\
&\geq&2(2m+1)^t-2\sum_{k=1}^{t}(2m+1)^{t-k}+1-\sum_{k=1}^{t}(2m+1)^{-k} \nonumber\\
&=&2(2m+1)^t-2\cdot \frac{(2m+1)^t-1}{2m}+1-(2m+1)^{-t}\frac{(2m+1)^t-1}{2m}\nonumber\\
&>&0
\end{eqnarray}
and
\begin{eqnarray}\label{DX_DY1}
&&D_{\{0, i_1, i_2, ..., i_{n_x} \}}- D_{\{0, j_1, j_2, ..., j_{n_y} \}} \nonumber\\
&=&(n_y-n_x)+\sum_{k=1}^{n_x}(2m+1)^{-i_k}-\sum_{k=1}^{n_y}(2m+1)^{-j_k} \nonumber\\
&\leq&-1+\sum_{k=1}^{t}(2m+1)^{-k} \nonumber\\
&=&-1+(2m+1)^{-t}\frac{(2m+1)^t-1}{2m}\nonumber\\
&<&0
\end{eqnarray}
 Case II. If  $n_x<n_y$, similar to case I, we have
\begin{equation}\label{TX_TY2}
T_{\{0, i_1, i_2, ..., i_{n_x} \}}- T_{\{0, j_1, j_2, ..., j_{n_y} \}} <0
\end{equation}
and
\begin{equation}\label{DX_DY2}
D_{\{0, i_1, i_2, ..., i_{n_x} \}}- D_{\{0, j_1, j_2, ..., j_{n_y} \}} >0
\end{equation}
Case III. If $n_x=n_y$,   Eq.(\ref{TX_TY})and (\ref{DX_DY}) can be rewritten as
\begin{eqnarray}\label{TX_TY3}
&&T_{\{0, i_1, i_2, ..., i_{n_x} \}}- T_{\{0, j_1, j_2, ..., j_{n_y} \}} \nonumber\\
&=&2\sum_{k=1}^{n_y}(2m+1)^{t-j_k}-2\sum_{k=1}^{n_x}(2m+1)^{t-i_k}+\sum_{k=1}^{n_y}(2m+1)^{-j_k}-\sum_{k=1}^{n_x}(2m+1)^{-i_k} \nonumber
\end{eqnarray}
and
\begin{equation}\label{DX_DY3}
D_{\{0, i_1, i_2, ..., i_{n_x} \}}- D_{\{0, j_1, j_2, ..., j_{n_y} \}}
=\sum_{k=1}^{n_x}(2m+1)^{-i_k}-\sum_{k=1}^{n_y}(2m+1)^{-j_k} \nonumber
\end{equation}
If $i_l=j_l$ holds for $l=1,2,\cdots, n_x$, it is easy to obtain
$$T_{\{0, i_1, i_2, ..., i_{n_x} \}}-T_{\{0, j_1, j_2, ..., j_{n_y} \}}=0$$
$$D_{\{0, i_1, i_2, ..., i_{n_x} \}}-D_{\{0, j_1, j_2, ..., j_{n_y} \}}=0$$
If there is a positive integer $k_0(1\leq k_0 \leq n_x)$, such that $i_l=j_l$ holds for $l=1,2,\cdots,k_0-1$, but $i_{k_0}\neq j_{k_0}$. It can be further divided into $2$ case.
\begin{itemize}
\item If  $i_{k_0}> j_{k_0}$, we have
 \begin{eqnarray}\label{TX_TY4}
 \hspace{-10mm}&&T_{\{0, i_1, i_2, ..., i_{n_x} \}}- T_{\{0, j_1, j_2, ..., j_{n_y} \}} \nonumber\\
 \hspace{-10mm}&=&2\sum_{k=1}^{n_y}(2m+1)^{t-j_k}-2\sum_{k=1}^{n_x}(2m+1)^{t-i_k}+\sum_{k=1}^{n_y}(2m+1)^{-j_k}-\sum_{k=1}^{n_x}(2m+1)^{-i_k} \nonumber\\
 \hspace{-10mm}&=&2\sum_{k=k_0}^{n_y}(2m+1)^{t-j_k}-2\sum_{k=k_0}^{n_x}(2m+1)^{t-i_k}+\sum_{k=k_0}^{n_y}(2m+1)^{-j_k}-\sum_{k=k_0}^{n_x}(2m+1)^{-i_k} \nonumber\\
 \hspace{-10mm}&\geq&2(2m+1)^{t-j_{k_0}}-2\sum_{k=j_{k_0}+1}^{t}(2m+1)^{t-k}+(2m+1)^{-j_{k_0}}-\sum_{k=j_{k_0}+1}^{t}(2m+1)^{-k} \nonumber\\
 \hspace{-10mm}&>&0
\end{eqnarray}
and
 \begin{eqnarray}\label{DX_DY4}
 &&D_{\{0, i_1, i_2, ..., i_{n_x} \}}- D_{\{0, j_1, j_2, ..., j_{n_y} \}} \nonumber\\
 &=&\sum_{k=1}^{n_x}(2m+1)^{-i_k}-\sum_{k=1}^{n_y}(2m+1)^{-j_k} \nonumber\\
 &=&\sum_{k=k_0}^{n_x}(2m+1)^{-i_k}-\sum_{k=k_0}^{n_y}(2m+1)^{-j_k} \nonumber\\
 &\leq &\sum_{k=j_{k_0}+1}^{t}(2m+1)^{-k}-(2m+1)^{-j_{k_0}} \nonumber\\
 &=&\frac{(2m+1)^{-j_{k_0}}-(2m+1)^{-t}}{2m}-(2m+1)^{-j_{k_0}} \nonumber\\
 &<&0
\end{eqnarray}
 \item If $i_{k_0}<j_{k_0}$, by symmetry, we have
 $$T_{\{0, i_1, i_2, ..., i_{n_x} \}}-T_{\{0, j_1, j_2, ..., j_{n_y} \}}<0$$
 $$D_{\{0, i_1, i_2, ..., i_{n_x} \}}-D_{\{0, j_1, j_2, ..., j_{n_y} \}}>0$$
  \end{itemize}






\end{document}